\documentstyle[11pt]{article}

\newcommand{\nn}{\nonumber}
\newcommand{\be}{\begin{equation}}
\newcommand{\ee}{\end{equation}}
\newcommand{\ba}{\begin{eqnarray}}
\newcommand{\ea}{\end{eqnarray}}
\newcommand{\half}{\frac{1}{2}}
\newcommand{\Tr}{\hbox{\rm Tr\ }}

\newcommand{\A}{{\cal A}}
\newcommand{\F}{{\cal F}}
\newcommand{\M}{{\cal M}}
\newcommand{\B}{{\cal B}}
\newcommand{\R}{{\cal R}}
\newcommand{\C}{{\cal C}}
\renewcommand{\O}{{\cal O}}
\renewcommand{\d}{{\rm d}}
\newcommand{\la}[1]{\label{#1}}
\newcommand{\nr}[1]{(\ref{#1})}
\begin{document}


\begin{flushright}
hep-th/9609108
\end{flushright}

\vskip 0.7truecm

\begin{center}
\large
{\bf Symmetries and observables for BF--theories in superspace }

\vskip 1.5cm

{\bf Pirjo Pasanen}\footnote{
 E-mail: PIRJO.PASANEN@HELSINKI.FI}
\\
\vskip 0.4cm
{\it Research Institute for Theoretical Physics \\
P.O. Box 9, FIN-00014 University of Helsinki, Finland}
\vskip0.2cm
\end{center}

\vskip 2cm
\noindent

The supersymmetric version of a topological quantum field theory
describing flat connections, the super BF-theory, is studied in the
superspace formalism.  A set of observables related to topological
invariants is derived from the  curvature of the superspace. Analogously to
the non-supersymmetric versions, we find that the theory exhibits a
vector-like supersymmetry.  The role of the vector supersymmetry and
an additional new symmetry of the action in the construction of
observables is explained.

\vskip 2cm

\section{Introduction}

Topological field theories offer an intriguing possibility to combine
ideas from physics and mathematics. They are quantum field theories
with no physical degrees of freedom and their properties are fully
determined by the global structure of the manifold they are defined
on. A remarkable feature is that for many topological theories, like
the Donaldson theory \cite{Wit-TQFT} and Chern-Simons (CS) theory,
 the expectation values of the
observables are topological invariants. 

The Chern-Simons  theory provides a
 three-dimensional interpretation of the theory of knots: the
 correlators of its observables, Wilson loops, are related to the
 Jones polynomials of knot theory \cite{Wit-Jones}. Another important
application of CS theory is $2+1$ dimensional gravity.
 CS action with Poincar\'e group as the gauge
group is the Einstein-Hilbert action \cite{2+1}, giving a gauge
theory interpretation of gravity in $2+1$ dimensions.
However, the
 Chern-Simons theory is defined only in three dimensions.  Its
 generalization to arbitrary dimensions \cite{KaRo-arb,Horo-ex} are
 called BF-models or antisymmetric tensor models. They, like the
 CS theory, describe the moduli space of flat connections
 and their observables are related to the linking and intersection
 numbers of manifolds.  The supersymmetric BF-theories (SBF) were
 introduced in \cite{Wit-topgra} as a supersymmetric version of $2+1$
 dimensional topological gravity. There it was also shown that the
 partition function of three dimensional SBF computes a topological
 invariant, the Casson invariant. Generalizations of SBF to other
 dimensions were considered in \cite{BBT} and \cite{Wal-alg,Horo-ex}.

In this article we study supersymmetric BF models. We are particularly
interested in finding new observables and possible topological
invariants for $3d$ SBF-theories, besides the partition function. By 
 formulating the theory in superspace a large set
of observables, including previously unknown ones, can be derived form
the superspace curvature. In \cite{BiRa-JGeom} a vector-like 
supersymmetry 
similar to that found in ordinary BF-models and Chern-Simons theory 
\cite{DGS-3d,BRT-ren}
was constructed for the SBF-models. In particular, the hierarchy of
observables constructed from the supercurvature can be derived from
one initial observable with the help of the vector
supersymmetry. Using the superspace formulation we extend this
construction to include also the anti-BRST and anti-vector
supersymmetries, in addition to the usual BRST and
vector supersymmetries. 

This article  is organized
as follows: in section 2 we will introduce the model and write
it in the superspace. It turns out that in the superspace formalism
many features of
the CS theory can be generalized directly to SBF-theory. In
section 3 we derive the set of observables and discuss their relation
to topological invariants. In section 4 we generalize the vector
supersymmetry to SBF and show how it can be used to construct
of new observables.

\section{Supersymmetric BF-theories}
\la{SBF}

The classical action or non-supersymmetric BF-models in $d$
dimensions is \be S_0 = \int \d^dx\, B_n^0 F_A, \nn \ee where $B_n^0$
is a $n=d-2$ form (with ghost number zero) and $F_A$ is the curvature
two form $F_A = \d A + \half [A,A]$.  In addition to the normal
Yang-Mills gauge symmetry $ A \to A + \d_A \omega_0$ the action is
invariant under the transformation $ B_n \to B_n+ \d_A\omega_{n-1}$
caused by the Bianchi identity. In dimensions higher than three this
symmetry is reducible:
$$
\omega_{n-1} \to \d_A\omega_{n-2} \quad\hbox{\rm etc. }
$$
 and additional ghost fields are needed in order to fix the gauge
according to the Batalin-Vilkovisky procedure.

In three dimensions the BF theory is closely related to the
Chern-Simons theory: CS-theory for the tangent group $TG \simeq (G,
\underline{g})$ is equivalent to the BF-theory for $G$
\cite{BBRT-TQFT}. In $TG$ the Chern-Simons connection one-form
splits into two parts $A$ and $B$, the basic fields of the BF-theory.
This makes it possible to construct the classical action of BF-theories, find
the BRST-transformations and fix the gauge easily by studying the CS
theory for the tangent group.

For the supersymmetric extension of three-dimensional BF-model the
 situation is quite similar --- the action and many properties of the
 theory can be expressed in terms of super CS-theory. This is done
 elegantly by formulating the theory in superspace with two
 anticommuting Grassmannian coordinates $\theta$, $\bar\theta$ in
 addition to the normal space time coordinates $x_\mu$. Here we will
 mainly concentrate in the three-dimensional case but with slight
 modifications the method is suited for SBF-models in other
 dimensions.

The integration over the Grassmannian variables is normalized as
\be
\int  \d\bar\theta\d\theta 
\left\{\begin{array}{c}
1 \\
\theta \\
\bar\theta \\
\theta\bar\theta
\end{array}\right\} = 
\left\{\begin{array}{c}
0\\
0\\
0\\
1
\end{array}\right\}. 
\ee
 If the coordinates  $\theta$ and $\bar \theta$are associated  with
ghost numbers $-1$ and 1 the superspace connection one-form $\hat\A$
in $3+2$ dimensions is written\footnote{
Note that we will use graded differential forms $X^q_p$ with ordinary
form degree $p$ and ghost number $q$. Two graded forms satisfy $X^q_p
Y^r_s = (-1)^{(q+p)(r+s)}Y^r_sX^q_p $. All the commutators should also
be considered  graded.}
\be
\hat{\A} = \hat{ A}^0_\mu\d x^\mu + \hat{A}^1_\theta \d\theta +
\hat{A}^{-1}_{\bar\theta}\d\bar\theta
\la{suconn}
\ee
where 
the superfields $\hat A^0_\mu\d x^\mu ,\ \hat A^1_\theta $ and 
$\hat A^{-1}_{\bar\theta}$ can be further expanded as:
\ba
\hat A^0_\mu    &=& A_\mu -\theta\psi_\mu + \bar\theta\chi_\mu +
\theta\bar\theta B_\mu\nn\\
\hat A^1_\theta    &=& c -\theta\phi + \bar\theta\rho +
\theta\bar\theta\eta\nn\\
\hat A^{-1}_{\bar\theta} &=& \bar\eta -\theta\bar\rho - \bar\theta\bar\phi
+ \theta\bar\theta \bar c.\nn
\ea
The components
 can be identified with the fields of three dimensional super
 BF-theory: $\psi^1_\mu$ and $\chi^{-1}_\mu$ are the superpartners of
 the connection $A^0_\mu$ and field $B^0_\mu$, while $ \rho^0_0,
 \bar\rho^0_0$ and $\phi^2_0, \bar\phi^{-2}_0$ are their corresponding
 ghosts and antighosts.
With these definitions  the classical action of the SBF-model can be
written as
the action of the super-CS theory: 
\be S_{cl} = \int \d^3x (BF_A -
\chi\d_A\psi)= \half\int\d^3x \d^2\theta (\hat\A \hat\d\hat\A +
\frac{2}{3} \hat\A[\hat\A,\hat\A] ) .  
\la{SBFcl} 
\ee 
To obtain the quantum action one has  to fix the gauge symmetry $\hat
A^0 \to \hat A^0 + \d_{\hat A^0} \omega$ by adding to the action a
BRST-exact gauge fixing term.

The BRST transformations of the fields can be derived from the 
superspace curvature two-form
using a method similar to that of
\cite{BiRa-JGeom,Wal-alg,MaiNie,HiNiTi,unicon,BiRaTh-red} for
Donaldson theory and
Witten type topological theories. But because of the $N=2$ superspace
with two anticommuting coordinates of opposite ghost numbers we can
extend this method to include also the anti-BRST symmetry. We 
define  the superspace curvature as
 \be
\hat\F = (\d x^\mu \partial_\mu + \d\theta  \delta +\d\bar\theta  
\bar\delta )
 \hat \A + \half [\hat\A, \hat \A ]
\la{scurv}
\ee
and impose the `horizontality condition'' 
\be
\hat\F \equiv \hat F_{\mu\nu}\d x^\mu \d x^\nu -
(\d\theta\partial_\theta + \d\bar\theta\partial_{\bar\theta})\hat \A ,
\la{restrict}
\ee
which  truncates the curvature to the physical part  independent of 
$\d\theta,\ \d\bar\theta$ (and consequently of the ghost fields),
 and identifies  the BRST-operator $\delta$
 with $\partial_\theta$ and $\bar\delta$ with
$\partial_{\bar\theta}$.  
This gives the BRST-transformations for the component fields:
\be
\begin{array}{llll}
\delta A          &=-\d_A c +\psi \quad
         &\delta B         &= -\d_A \eta - [c,B] + [\phi,\chi]+
[\psi,\rho]\cr
\delta c &= -\half [c,c] +\phi \quad  
         &\delta \eta      &=-[c,\eta] +[\phi,\rho] \cr
\delta \psi       &=  -\d_A \phi - [c,\psi] \quad 
         &\delta \chi      &= -\d_A\rho - [c,\chi] +B\cr
\delta \phi       &=  -[c,\phi]  \quad
         &\delta \rho      &= -[c,\rho] +\eta 
\end{array}
\la{BRST1}
\ee  
which have to be supplemented with the transformations of the
anti-ghosts and Lagrange multipliers $\lambda_0^0, b_0^0,
\beta_0^{-1}, \sigma_0^1$ for the gauge fixing conditions of the 
fields $A,B,\psi$ and $\chi$. The Lagrange multipliers 
can be combined into a superfield
\be 
\hat\Lambda_0^0 = \lambda -\theta \sigma
-\bar\theta\beta +\theta\bar\theta b .
\la{lambda} 
\ee  
The simplest choice for the BRST transformations would be
$$
\delta \hat\A^{-1}_0 = -\hat\Lambda, \qquad\delta \hat\Lambda=0
$$
but with suitable field redefinitions 
these can be put into a form  which will be more convenient later on:
\be
\begin{array}{llll}
\delta \bar c     &=  -b \quad
         &\delta \bar\eta&= -\lambda -[c,\bar\eta] + \bar\rho\\
\delta b &= 0\quad
          &\delta\lambda &= -[c,\lambda] -[\phi,\bar\eta]
-\sigma \\
\delta \bar\phi &= \beta - \bar c  \quad 
         &\delta \bar\rho&= \sigma -[c,\bar\rho] \\
\delta \beta      &=    -b \quad
         &\delta \sigma    &= -[c,\sigma] +[\phi,\bar\rho].
\end{array}
\la{BRST2}
\ee
The gauge fixing part of the supersymmetric 
action is chosen to be 
\ba
S_{gf} &=& \int \d^2\theta\, 
	\delta(\hat A^{-1}_{\bar\theta} \d * \hat A^0 )\cr
&=&\int \d^3x \,(-b \d *A -\lambda \d *B 
	+ \beta\d *\psi +  \sigma\d *\chi 
	+ \bar c \d *\d_A  c +\bar \eta \d *\d_A \eta \la{SBFgf}\\
&+&  \bar \phi\d *\d_A \phi + \bar \rho \d *\d_A \rho 
        - \bar\eta [\d c, *B] + \bar\eta \d [\phi, *\chi] 
	- \bar\eta \d [\rho, *\psi]+ \bar\rho [\d c, *\chi] 
 +\bar\phi \d [c, *\psi]) 
\nn
\ea
Note also that unlike in the ordinary BF-model the classical action 
\nr{SBFcl}
is now BRST-exact
\[
S_{cl} = \int \d^3x\, \delta(\chi F_A ).
\]
This shows that the supersymmetric BF model is a Witten type
topological theory with a $\delta$-exact action, whereas the ordinary
non-abelian BF models are Schwartz type theories  \cite{BBRT-TQFT}.

The gauge fixing term $S_{gf}= \int \d^2\theta\, 
	\delta(\hat A^{-1}_{\bar\theta} \d * \hat A^0)$ is formally
similar to that of Chern-Simons theory quantized in Landau gauge $\d
*\A =0$:
\[
S^{CS}_{gf}= \int \d^3 x (\delta \bar \C \d * \A)
\]
In CS theory the BRST and anti-BRST operators are related by
transformation obtained by integrating the quantum action by parts 
\cite{BiRa-vecSUSY}
\[
S^{CS}_q = \int \d^3 x (\A \d\A +
\frac{2}{3} \A[\A,\A] - \Lambda \d * \A - \bar \C \d *\d_\A \C).
\]
The
integrated action is equivalent to the original action after a change of
 fields which leaves the connection $\A$ unchanged but takes the
ghosts $\C$ to the antighosts $\bar\C$ and $\bar\C$ to $-\C$. The
Lagrange multiplier field $\Lambda$ transforms as $\Lambda \to \Lambda
- [\C, \bar\C]$. This transformation of the fields maps $\delta$ to
$\bar\delta$ .

For the super-CS and consequently for the three dimensional
SBF-theory the situation is again analogous.  Integrating the gauge
fixed quantum action $S_{q} = S_{cl}+ S_{gf}$
(\ref{SBFcl},\ref{SBFgf}) by parts we find the superspace version of
the transformation which relates BRST- and anti-BRST operators. In
superspace language the transformation rules can be expressed
compactly by demanding that under the ``conjugation'' of the Grassmann
variables 
\be
\theta \to \bar \theta,\qquad  \bar\theta \to - \theta
\la{conjugation}
\ee
the total superspace connection $\hat\A$ stays the same while the
operators change as $\delta\to\bar\delta,\  \bar\delta\to
-\delta$. The transformations for the Lagrange multipliers are somewhat
more complicated
\be
\begin{array}{llll}
\lambda &\to \lambda + [c, \bar\eta]\qquad  
& b &\to b -[c, \bar c]-[\eta , \bar \eta ]
         -[\rho , \bar\rho ]-[\phi ,\bar\phi] \cr
\sigma &\to \beta + [c, \bar\phi] \qquad &\beta &\to -\sigma
+[\phi,\bar\eta].
\la{trans}
\end{array} 
\nn
\ee

For BF-theories in dimensions other than three the situation is
slightly more complicated because  the $A$ and $B$
fields cannot be combined 
into one connection. In $d$ dimensions $B$ is a $d-2$ form
and  additional fields will be needed to take care of the reducibility. 
It is however possible to use truncated fields and write the
components of $\hat\A$ as
\ba
\hat A_\mu^0 &=&  A_\mu - \theta \psi_\mu \cr
\hat A_\theta^1 &=&  c - \theta \phi \cr
\hat A_{\bar\theta}^{-1} &=& -\bar\theta\bar\phi  + \theta\bar\theta \bar 
c 
\la{trunc}
\ea
and similarly for the $d-2$ superform $\hat\B$, which now contains in
addition to $B$, $\chi$ and their ghosts also the whole tower of
ghosts for ghosts from the Batalin-Vilkovisky gauge fixing. The
curvature of the $B$ sector is defined as 
\be \hat\R =(\d x^\mu
\partial_\mu + \d\theta\delta )\hat\B + [\hat\A , \hat\B ].  
\ee
 It
satisfies a Bianchi identity, and again after imposing the
horizontality condition similar to \nr{restrict} it reproduces the
correct nilpotent BRST-transformations. Since the $A$ and $B$ sectors
do not appear symmetrically in the action there exists no partial
integration symmetry and thus no anti-BRST operator $\bar\delta$.

\section{Observables}
\la{Obs}
 
In order to establish that the observables of the theory are indeed
topological invariants it must be checked that they are BRST-closed,
their expectation values do not depend on variations of the metric
and, if they are integrals of some local functionals, that their
BRST-cohomology depends only on homology class of the integration
contour. The partition function of the three dimensional SBF-model 
$$
Z_{3d} = \int e^{iS_q}.
$$
obviously satisfies all the requirements and  can be shown to equal
 the Casson invariant of the manifold \cite{Wit-topgra,BlaTho-Casson}.
 We will now derive a set of other observables for $3d$ SBF form the
 superspace curvature \nr{scurv} and see if they too could correspond
 to topological invariants.

The Bianchi identity 
\be
(\d x^\mu \partial_\mu +\d\theta\delta + \d\bar\theta\bar\delta)\hat\F
 + [\hat\A,\hat\F] =0
\nn
\ee
 guarantees that the powers of $\hat\F$ obey
\be
(\d x^\mu \partial_\mu +\d\theta\delta + \d\bar\theta\bar\delta ) 
\Tr {\hat \F}^n =0.
\la{trcurv}
\ee
The simplest one is the superspace 4-form ${\hat\F}^2$. It can be
expanded in powers of $\d\theta$ and $\d\bar\theta$:
\be
\half \Tr {\hat\F}^2 = \sum_{i,j; i+j\le 4}
 W^{i,j} {\d\theta}^i{\d\bar\theta}^j .
\nn
\ee
Equation \nr{trcurv} gives
\be
\d W^{i,j} + \delta W^{i-1,j} + \bar\delta W^{i,j-1} = 0.
\la{cohom}
\ee
When $j=0$ the integrals of the $(4-i)$-form $W^{i,0}$ over a
$(4-i)$ cocycle $\gamma$ are BRST-closed:
\be
\int_\gamma \d W^{i,0} +  \delta \int_\gamma W^{i-1,0} =
\int_{\partial\gamma} W^{i,0} + \delta \int_\gamma W^{i-1,0} =
\delta\int_\gamma W^{i-1,0} =0. \nn
\ee
Because of \nr{cohom} the BRST-cohomology of $\int W$ depends only on the
homology class of $\gamma$, making the vacuum expectation values and
correlation functions of $\int W$ good candidates for topological
invariants.
Note that because of the symmetry of the three-dimensional action
(\ref{SBFcl},\ref{SBFgf}) under the
partial integration transformation the expectation values of  any
observable $\O $ and its ``conjugate''
$\overline{\O}$ 
are the same. This can be seen by making a change of variables (with a unit
Jacobian) in the path integral taking all the fields to their
conjugates and using the invariance of the action. The condition
$\delta \O =0$ changes under this transformation to $\bar\delta
\overline{ \O} =0$. Therefore objects that are either $\delta$- or
$\bar\delta$-closed qualify as observables of $3d$ SBF. In particular,
we can thus identify $\overline{W}^{i,j} = (-1)^j W^{j,i}$.

The expansion of ${\hat\F}^2$ gives using \nr{restrict} 
\ba
W^{00} &=& \half F^2  \cr
W^{10} &=& \psi F - \bar\theta (BF - \chi \d_A \psi)\cr
W^{20} &=& \half \psi^2 +\phi F + \theta (\phi\d_A \psi) 
         - \bar\theta (\psi B + \phi\d_A\chi + F\eta)\cr
       &+& \theta\bar\theta (
         \phi \d_A B - \phi [\psi,\chi] + \d_A \psi \eta) \cr
W^{30} &=& \psi\phi -\bar\theta (\phi B + \psi\eta) \cr
W^{40} &=& \half \phi^2 -\bar\theta (\phi\eta) 
\la{W}
\ea
{}from which we can extract 11 observables. The previously unknown ones
are the $\theta$- and
$\theta\bar\theta$-components of $W^{20}$. They  are a particular to 
 three-dimensional theories and unlike the others
 which, or rather their generalizations involving
all the Batalin-Vilkovisky ghosts, can be obtained from the truncated
supercurvatures $\hat\F$ and $ \hat\R$ of $A$ and $B$-sectors in
all dimensions. Nevertheless, the $\theta$ component of $W^{20}$
 seems to be BRST-closed also in higher dimensions:
the ghosts for ghosts and other fields appear only in the
transformations for the $B$-sector.

Interestingly, some of the
observables above are formally the same as for Donaldson theory. This is no
surprise since the BRST-structure of Donaldson theory is similar to
that of the $A$-sector of the SBF. In fact, the SBF can be thought as
reduction of the Donaldson theory to three dimensions
\cite{BiRaTh-red,MaiNie}.

As characteristic for the Witten type topological theories the
expectation values of observables
\be
<\O> = \int [dX] \O e^{i/g^2 S_q}
\nn
\ee
are independent of the coupling $g^2$. The integral can  be calculated
 in the $g^2 \to 0$ limit where it localizes to
the classical equations of motion
\be
F_A=0, \qquad \d_A \psi =0, \qquad \d_A B - [\psi,\chi]=0,\qquad
\d_A\chi =0 
\la{EQM}
\ee
{\it i.e.\, }it is now calculated over the moduli space of flat
connections $\M$. In the  limit $g^2 \to 0$ the fields in \nr{W} are
replaced by their classical values \nr{EQM}. Then the non-vanishing
observables are
\be
\begin{array}{llllll}
\omega_0^4 &= \half\phi^2\quad  &\omega_1^3 &= \int\psi\phi \quad
&\omega_2^2 &= \half\int \psi^2\\
\omega_0^3 &= \phi\eta \quad &\omega_1^2 &=\int \psi\eta +
B\phi\quad 
&\omega_2^1 &= \int\psi B.
\end{array}
\nn
\ee  

To evaluate the expectation values one has to take care of the zero
modes of the fermions. Especially in dimensions higher than two the
zero-modes of the other fields complicate matters considerably. We
will not perform the calculations here but refer to 
\cite{BBRT-TQFT} and references therein for
discussions on topological invariants of the Donaldson theory. The
considerations there are quite similar to those for the observables
$\omega_0^4,\ \omega_1^3,\ \omega_2^2$ of the $A$-sector of SBF.  The
invariant corresponding to $\omega^2_2$ has been evaluated in
\cite{BBRT-TQFT} for $2d$ BF and found to be the symplectic volume of
the moduli space. Its products with $\omega^4_0$ produce linking and
intersection numbers of moduli spaces.

\section{Vector supersymmetry and the tower of observables}
\la{VecSUSY}

A peculiar feature of Chern-Simons and BF-theories is a vector-like
supersymmetry of the action \cite{BRT-ren}. It gives rise to new Ward
identities which have been utilized in proving the theories to be
finite, renormalizable and free of anomalies \cite{vecSUSY-1,
vecSUSY-2}. However, this supersymmetry is valid only in flat
space. One might argue though that because the theories are
topological their physical quantities are not dependent on the metric
of the manifold. In any case, the vector supersymmetry has been
established as a common feature of many topological theories
\cite{BiRa-JGeom} and a
useful tool, not only in the study of renormalization and related
topics but also in finding new observables.

The vector supersymmetry for non-supersymmetric CS theory quantized
in Landau gauge is generated by operator $s$ with ghost number and
form degree 1:
\be
\begin{array}{llll}
s \A &= *\d\C \qquad &s \C &= 0 \nn\\
s \bar \C &= \A \qquad  &s \Lambda &= -\delta \A,
\end{array}
\la{vectorSUSY}
\ee
or written in component form
\[
s = s_\alpha\d x^\alpha \qquad
*\d \C =-\epsilon_{\mu\alpha\beta} \partial^\beta\C \d x^\mu\d
x^\alpha.
\]
  Using the partial integration for the CS theory one can obtain
 the anti-supersymmetry $\bar s$:
\be
\begin{array}{llll}
\bar s \A &= *\d\bar\C \qquad & \bar s  \bar\C &= 0 \nn\\
\bar s \C &= -\A \qquad  & \bar s \Lambda &= - \bar\delta \A -
[\A,\bar \C]
\end{array}
\la{avectorSUSY}
\ee
 The anti\-commutation relations of the operators $\delta,\bar\delta$
and $s, \bar s$ are
\ba
[s_\alpha , s_\beta ]&=&[\bar s_\alpha , \bar s_\beta ]=
[\delta, \bar\delta ] = [s_\alpha ,\bar s_\beta ] 
= [\delta , s_\alpha ] = [\bar\delta, \bar s_\alpha ] =0 
\la{commu}
\\
{} [\delta, \bar s_\alpha ] &=& -[\bar\delta , s_\alpha ] 
         = \partial_\alpha + 
\hbox{\rm \, terms vanishing modulo the equations of motion}.
\nn
\ea
Together with the BRST-operators $\delta$ and $\bar\delta$ the
operators $s$ and $\bar s$ can be combined to form a generator of
$N=2$ supersymmetry algebra \cite{DGS-3d,BiRa-vecSUSY,BiRa-JGeom}.  The 
vector
supersymmetries can be formulated also for the non-supersymmetric
BF-theories. In dimensions other than three there exists no vector
supersymmetry $s$ but the $\bar s$ operator can still be constructed
\cite{vecSUSY-1}.

The anti-vector supersymmetry can be generalized to the
supersymmetric BF-theory in arbitrary dimensions.  In $3d$ it can be
derived easily using \nr{avectorSUSY} for the superfields $\hat A^0,
\hat A^1_\theta, \hat A^{-1}_{\bar\theta}, \hat\Lambda$:
\be
\begin{array}{llllll}
\bar s_\alpha A_\mu
&=&-\epsilon_{\mu\alpha\beta}\partial^\beta\bar\eta\quad
         &\bar s_\alpha B 
         &=& -\epsilon_{\mu\alpha\beta}\partial^\beta\bar c \cr
\bar s_\alpha c &=& A_\alpha\quad
         &\bar s_\alpha  \eta &=& B_\alpha\cr
\bar s_\alpha \bar c &=&0 \quad
         &\bar s_\alpha \bar\eta &=&0 \cr
\bar s_\alpha b &=& -\partial_\alpha \bar c\quad
         &\bar s_\alpha \lambda &=& D_\alpha \bar\eta\cr
\bar s_\alpha \psi_\mu  
&=& \epsilon_{\mu\alpha\beta}\partial^\beta\bar\rho  \quad 
         &\bar s_\alpha \chi 
         &=&\epsilon_{\mu\alpha\beta}\partial^\beta\bar\phi \cr
\bar s_\alpha \phi &=& -\psi_\alpha\quad
         &\bar s_\alpha \rho &=& -\chi_\alpha \cr
\bar s_\alpha \bar \phi &=&0 \quad
         &\bar s_\alpha \bar\rho &=&0 \cr
\bar s_\alpha \beta &=& \partial_\alpha \bar \phi\quad
         &\bar s_\alpha \sigma &=& D_\alpha \bar \rho 
\end{array}
\la{aSUSY}
\ee
This is a symmetry of the quantum action (\ref{SBFcl},\ref{SBFgf}) and
satisfies the anticommutation relations \nr{commu} with the
BRST-operator (\ref{BRST1},\ref{BRST2}).  The analysis done on the
renormalization, finiteness and anomalies of ordinary BF-theories
using vector supersymmetry can thus be applied directly to the
supersymmetric BF-theories.

It is interesting to note that the vector supersymmetry of the SBF can
be useful also in constructing new observables (see also
\cite{BiRa-JGeom} for a slightly different approach).  Whenever there exists
a BRST-closed object $\omega$ also $\bar s\omega$ is BRST-closed as
a result of the anticommutation relations \nr{commu}. So in principle
it is possible to find an observable, like $\omega_0^4$ and
$\omega_0^3$, and apply $\bar s_\alpha$ successively to get new
ones. Also, since 
\be
\omega_0^4 = \half\delta  (c\phi -\frac{1}{6}c[c,c]),\qquad 
\omega_0^3 = \half\delta ( \phi \rho +c\eta -\half \rho [c,c] )=
\delta (\phi\rho)
\nn
\ee
all observables obtained by acting with $\bar s$ are in fact
BRST-exact --- modulo equations of motion and surface terms. This is
valid only locally and does not mean that the observables should be
trivial.

{}From \nr{W} we see that modifying slightly the anti-supersymmetry
transformations for $\psi$ and $B$ in \nr{aSUSY} as 
\[
\bar s B = *\d \bar c -2\d_A \chi, \qquad \bar s\psi = -*\d\bar\rho
-2 F
\]
 and leaving the others intact the anti-supersymmetry still remains a
symmetry of the action.  By denoting  the metric independent part of
the modified $\bar s$ operator by $\bar v$: 
\be
\begin{array}{llllll}
\bar v B  &= -2\d_A\chi \quad &\bar v\eta &= -B\quad &\bar v\rho &= -\chi \cr
\bar v\psi&= -2F        \quad &\bar v c   &= -A\quad &\bar v\phi &= -\psi
\end{array}
\la{mods}
\ee 
and applying successively $\frac{1}{k!}(-\bar v)^k$ to $\half\phi^2$ and
 $\phi\eta$ 
 it is possible to derive all the observables in \nr{W},
except the $\theta$ and $\theta\bar\theta$ components of $W^{02}$.
 The symmetry   $\bar v$
acts as a vertical (in the direction of the form
 degree) transformation along the components $W^{ij}$ of the
 supersurvature \nr{scurv}.

It is easily seen that also a horizontal (ghost number direction)
transformation $\bar h$ can be defined:
\be
\begin{array}{llll}
\bar h A    &=\chi \quad &\bar h c   &= -\rho\cr
\bar h \psi &= -B  \quad &\bar h\phi &= -\eta\cr
\bar h\bar\eta &=-\bar\phi\quad &\bar h\bar\rho &= -\bar c .
\end{array}
\la{horizontal}
\ee
This is a symmetry of the action and commutes with the
BRST-operator. It thus allows us to construct all possible
observables starting from the element $\half \phi^2$ of highest ghost
number and lowest form degree --- again excluding the $\theta$ and
$\theta\bar\theta$ components of $W^{20}$.

The vertical transformation has geometrical interpretation as the
equivariant derivative of BRST-model acting on the curvature of the
universal bundle over the space of gauge connections
\cite{BiRa-uni}, which can be identified with the
supercurvature $\hat F$. Therefore the vertical transformation can be
defined for all Witten type topological theories. The horizontal
transformation $\bar h$ which can be constructed only in three
dimensions is in fact part of the anti-BRST operator
$-\delta$:  only those terms that are not composites of
 fields and  do not contain Lagrange multipliers are included.

Acting on
 $\phi^2$ with the total
 vector supersymmetry transformation $\bar s$ instead of the vertical 
transformation we get even a larger set
 of observables. In addition to theose present in \nr{W} these include
observables that depend explicitely on the metric.
 Since we already know that the observables in \nr{W}
 are BRST-closed also the metric dependent ones  should be closed
 separately. Moreover, it can be shown that the metric
 variations of these observables can be written as BRST-exact terms,
 a necessary requirement for the observables to be topological
 invariants \cite{BBRT-TQFT}. So we can conclude that the expectation
 values of these observables are of topological nature.

\section{Conclusions}

We have studied three dimensional supersymmetric BF-theories using
superspace formalism. This proved out to be a powerful method for
studying the properties of the theory and especially  for finding new
symmetries and observables. The superspace curvature gives rise to a
hierarchy of observables, which could be derived starting from one initial
observable using the two transformations we constructed.
 The transformations have a geometrical interpretation as
vertical and horizontal transformations  acting on the components of
the supercurvature, and can  be identified as parts of
more general symmetries of the action, the vector supersymmetry
and anti-BRST symmetry.

\subsection*{Acknowledgements}

We thank prof. A. J. Niemi for useful comments on the manuscript and
the referee for pointing out \cite{BiRa-JGeom}.


\end{document}